\documentclass[10pt]{article}
\usepackage{latexsym,graphicx}
\newcommand{\be}{\begin{equation}}
\newcommand{\ee}{\end{equation}}
\def\n{\noindent}
\catcode `\@=11
\catcode `\@=12
\begin{document}
\begin{center}
\large{\bf {Magnetized String Cosmological Model in Cylindrically Symmetric 
Inhomogeneous Universe with Variable Cosmological-Term $\Lambda$ }} \\
\vspace{10mm}
\normalsize{Anirudh Pradhan}\\
\vspace{5mm}
\normalsize{Department of Mathematics, Hindu Post-graduate College, 
Zamania-232 331, Ghazipur, India \\
E-mail : pradhan@iucaa.ernet.in, acpradhan@yahoo.com }\\
%\vspace{5mm}
%\normalsize{} 
%\vspace{5mm}
%\normalsize{}
\end{center}
\vspace{10mm}
%\date{}
%\maketitle
\begin{abstract} 
Cylindrically symmetric inhomogeneous magnetized string cosmological model is investigated with 
cosmological term $\Lambda$ varying with time. To get the deterministic solution, it has been 
assumed that the expansion ($\theta$) in the model is proportional to the eigen value 
$\sigma^{1}~~_{1}$ of the shear tensor $\sigma^{i}~~_{j}$. The value of cosmological constant for 
the model is found to be small and positive which is supported by the results from recent 
supernovae Ia observations. The physical and geometric properties of the model are also 
discussed in presence and absence of magnetic field.   
\end{abstract}
\smallskip
\n Keywords : String, Inhomogeneous universe, Cylindrical symmetry, Variable cosmological term \\
\n PACS number: 98.80.Cq, 04.20.-q, 98.80.Es 
%\newpage
%%%%%%%%%%%%%%%%%%%%%%%%%%%%%%%%%%%%%%%%%%%%%%%%%%%%%%%%%%%%%%%%%%%%%%%%%%%%%%%%%%%
%%%%%%%%%%%%%%%%%%%%%%%%%%%%%%%   SECTION 1  %%%%%%%%%%%%%%%%%%%%%%%%%%%%%%%%%%%%%%
\section{Introduction}
Cosmic strings play an important role in the study of the early universe. These strings 
arise during the phase transition after the big bang explosion as the temperature goes 
down below some critical temperature as predicted by grand unified theories 
\cite{ref1}${-}$ \cite{ref5}. It is believed that cosmic strings give rise to density 
perturbations which lead to formation of galaxies \cite{ref6}. These cosmic strings have 
stress energy and couple to the gravitational field. Therefore, it is interesting to study 
the gravitational effect which arises from strings. The general treatment of strings was 
initiated by Letelier \cite{ref7,ref8} and Stachel \cite{ref9}. The occurrence of magnetic 
fields on galactic scale is well-established fact today, and their importance for a variety 
of astrophysical phenomena is generally acknowledged as pointed out Zel'dovich \cite{ref10}. 
Also Harrison \cite{ref11} has suggested that magnetic field could have a cosmological origin. 
As a natural consequences, we should include magnetic fields in the energy-momentum tensor of the 
early universe. The choice of anisotropic cosmological models in Einstein system of field 
equations leads to the cosmological models more general than Robertson-Walker model \cite{ref12}. 
The presence of primordial magnetic fields in the early stages of the evolution of the universe 
has been discussed by several authors (Misner, Thorne and Wheeler \cite{ref13}; Asseo and Sol 
\cite{ref14}; Pudritz and Silk \cite{ref15}; Kim, Tribble, and Kronberg \cite{ref16}; Perley and 
Taylor \cite{ref17}; Kronberg, Perry and Zukowski \cite{ref18}; Wolfe, Lanzetta and Oren \cite{ref19}; 
Kulsrud, Cen, Ostriker and Ryu \cite{ref20}; Barrow \cite{ref21}). Melvin \cite{ref22}, in his 
cosmological solution for dust and electromagnetic field suggested that during the evolution 
of the universe, the matter was in a highly ionized state and was smoothly coupled with the field, 
subsequently forming neutral matter as a result of universe expansion. Hence the 
presence of magnetic field in string dust universe is not unrealistic. \\

Benerjee et al. \cite{ref23} have investigated an axially symmetric Bianchi type I string dust 
cosmological model in presence and absence of magnetic field. The string cosmological models with a magnetic 
field are also discussed by Chakraborty \cite{ref24}, Tikekar and Patel \cite{ref25,ref26}. Patel 
and Maharaj \cite{ref27} investigated stationary rotating world model with magnetic field. Ram 
and Singh \cite{ref28} obtained some new exact solution of string cosmology with and without a 
source free magnetic field for Bianchi type I space-time in the different basic form 
considered by Carminati and McIntosh \cite{ref29}. Singh and Singh \cite{ref30} investigated string 
cosmological models with magnetic field in the context of space-time with $G_{3}$ 
symmetry. Singh \cite{ref31} has studied string cosmology with electromagnetic fields in 
Bianchi type-II, -VIII and -IX space-times. Lidsey, Wands and Copeland \cite{ref32} have 
reviewed aspects of super string cosmology with the emphasis on the cosmological 
implications of duality symmetries in the theory. Bali et al. \cite{ref33,ref34,ref35} have 
investigated Bianchi type I magnetized string cosmological models.\\

Cylindrically symmetric space-time play an important role in the study of the universe 
on a scale in which anisotropy and inhomogeneity are not ignored. Inhomogeneous 
cylindrically symmetric cosmological models have significant contribution in 
understanding some essential features of the universe such as the formation of 
galaxies during the early stages of their evolution. Bali and Tyagi \cite{ref36} and 
Pradhan et al. \cite{ref37,ref38} have investigated cylindrically symmetric inhomogeneous 
cosmological models in presence of electromagnetic field. Barrow and Kunze \cite{ref39,ref40} 
found a wide class of exact cylindrically symmetric flat and open inhomogeneous string 
universes. In their solutions all physical quantities depend on at most one space 
coordinate and the time. The case of cylindrical symmetry is natural because of the 
mathematical simplicity of the field equations whenever there exists a direction in 
which the pressure equal to energy density. \\

In modern cosmological theories, a dynamic cosmological term $\Lambda(t)$ 
remains a focal point of interest as it solves the cosmological constant 
problem in a natural way. There are significant observational evidence for 
the detection of Einstein's cosmological constant, $\Lambda$ or a component 
of material content of the universe that varies slowly with time and space 
to act like $\Lambda$. A wide range of observations now compellingly suggest 
that the universe possesses a non-zero cosmological term \cite{ref41}. 
In the context of quantum field theory, a cosmological term corresponds to the 
energy density of vacuum. The birth of the universe has been attributed to an 
excited vacuum fluctuation triggering off an inflationary expansion followed 
by the super-cooling. The release of locked up vacuum energy results in 
subsequent reheating. The cosmological term, which is measure 
of the energy of empty space, provides a repulsive force opposing the 
gravitational pull between the galaxies. If the cosmological term exists, 
the energy it represents counts as mass because  mass and energy are equivalent. 
If the cosmological term is large enough, its energy plus the matter in the 
universe could lead to inflation. Unlike standard inflation, a universe with 
a cosmological term would expand faster with time because of the push from 
the cosmological term \cite{ref42}. Some of the recent discussions on the 
cosmological constant ``problem'' and on cosmology with a time-varying
cosmological constant by Ratra and Peebles \cite{ref43}, Dolgov \cite{ref44} 
and Sahni and Starobinsky \cite{ref45} point out that in the absence of any 
interaction with matter or radiation, the cosmological constant remains a 
``constant''. However, in the presence of interactions with matter or radiation, 
a solution of Einstein equations and the assumed equation of covariant conservation 
of stress-energy with a time-varying $\Lambda$ can be found. This entails that 
energy has to be conserved by a decrease in the energy density of the vacuum 
component followed by a corresponding increase in the energy density of matter 
or radiation (see also Weinberg \cite{ref46}, Carroll, Press and Turner \cite{ref47}, 
Peebles \cite{ref48}, Padmanabhan \cite{ref49} and Pradhan et al. \cite{ref50} ). \\

Recent observations by Perlmutter et al. \cite{ref51} and 
Riess et al. \cite{ref52} strongly favour a significant and a positive 
value of $\Lambda$ with magnitude $\Lambda(G\hbar/c^{3}) \approx 10^{-123}$. 
Their study is based on more than $50$ type Ia supernovae with 
red-shifts in the range $0.10 \leq z \leq 0.83$ and these suggest Friedmann 
models with negative pressure matter such as a cosmological constant 
$(\Lambda)$, domain walls or cosmic strings (Vilenkin \cite{ref53}, Garnavich 
et al. \cite{ref54}). Recently, Carmeli and Kuzmenko \cite{ref55} have shown 
that the cosmological relativistic theory predicts the value for cosmological 
constant $\Lambda = 1.934\times 10^{-35} s^{-2}$. This value of ``$\Lambda$'' 
is in excellent agreement with the recent estimates of the High-Z Supernova Team 
and Supernova Cosmological Project (Garnavich et al. \cite{ref54}; 
Perlmutter et al. \cite{ref51}; Riess et al. \cite{ref52}; 
Schmidt et al. \cite{ref56}). In Ref. \cite{ref57} Riess et al. have 
recently presented an analysis of 156 SNe including a few at $z > 1.3$ from the 
Hubble Space Telescope (HST) ``GOOD ACS'' Treasury survey. They conclude to the 
evidence for present acceleration $q_{0} < 0$ $(q_{0} \approx -0.7)$. Observations 
(Knop et al. \cite{ref58}; Riess et al., \cite{ref57}) of Type Ia Supernovae (SNe) 
allow us to probe the expansion history of the universe leading to the conclusion 
that the expansion of the universe is accelerating.\\

Recently Baysal et al. \cite{ref59} have investigated some string cosmological models in 
cylindrically symmetric inhomogeneous universe. Motivated by the situation discussed 
above, in this paper, we have generalized these solutions by including electromagnetic 
field tensor, pressure and cosmological term varying with time. We have taken strings and 
electromagnetic field together as the source gravitational field as magnetic field are 
anisotropic stress source and low strings are one of anisotropic stress source as well. 
The paper is organized as follows. The metric and the field equations are presented in Section 2. 
In Section 3, we deal with the solution of the field equations in presence of perfect fluid with 
electromagnetic field and variable cosmological term. Section 4 describes some physical and 
geometric properties of the universe. Finally in Section 5 concluding remarks are given. 
%%%%%%%%%%%%%%%%%%%%%%%%%%%%%%%%%%%%%%%%%%%%%%%%%%%%%%%%%%%%%%%%%%%%%%%%%%%%%%%%%%%
%%%%%%%%%%%%%%%%%%%%%%%%%%%%%%%  SECTION 2  %%%%%%%%%%%%%%%%%%%%%%%%%%%%%%%%%%%%%%%%
\section{The Metric and Field  Equations}
We consider the metric in the form 
\begin{equation}
\label{eq1}
ds^{2} = A^{2}(dx^{2} - dt^{2}) + B^{2} dy^{2} + C^{2} dz^{2},
\end{equation}
where $A$, $B$ and $C$ are functions of $x$ and $t$.
The energy momentum tensor for the cloud of strings with perfect fluid and electromagnetic field 
has the form 
\begin{equation}
\label{eq2}
T^{j}_{i} = (\rho + p)u_{i}u^{j} + p g^{j}_{i} - \lambda x_{i}x^{j} +  E^{j}_{i},
\end{equation}
where $u_{i}$ and $x_{i}$ satisfy conditions
\begin{equation}
\label{eq3}
u^{i} u_{i} = - x^{i} x_{i} = -1,
\end{equation}
and
\begin{equation}
\label{eq4}
u^{i} x_{i} = 0.
\end{equation}
Here $\rho$ is the rest energy density of the cloud of strings, $p$ is the isotropic pressure, 
$\lambda$ is the tension density of the strings, $x^{i}$ is a unit space-like vector representing 
the direction of strings so that $x^{1} = 0 = x^{2} = x^{4}$ and $x^{3} \ne 0$, 
and $u^{i}$ is the four velocity vector satisfying the 
following conditions
\begin{equation}
\label{eq5}
g_{ij} u^{i} u^{j} = -1.
\end{equation}
In Eq. (\ref{eq2}), $E^{j}_{i}$ is the electromagnetic field given by Lichnerowicz \cite{ref60} 
\begin{equation}
\label{eq6}
E^{j}_{i} = \bar{\mu}\left[h_{l}h^{l}\left(u_{i}u^{j} + \frac{1}{2}g^{j}_{i}\right) 
- h_{i}h^{j}\right],
\end{equation}
where $\bar{\mu}$ is the magnetic permeability and $h_{i}$ the magnetic flux vector 
defined by
\begin{equation}
\label{eq7}
h_{i} = \frac{1}{\bar{\mu}} \, {^*}F_{ji} u^{j},
\end{equation}
where the dual electromagnetic field tensor $^{*}F_{ij}$ is defined by Synge \cite{ref61} 
\begin{equation}
\label{eq8}
^{*}F_{ij} = \frac{\sqrt{-g}}{2} \epsilon_{ijkl} F^{kl}.
\end{equation}
Here $F_{ij}$ is the electromagnetic field tensor and $\epsilon_{ijkl}$ is the Levi-Civita 
tensor density.

In the present scenario, the comoving coordinates are taken as 
\begin{equation}
\label{eq9}
u^{i} = \left(0, 0, 0, \frac{1}{A}\right). 
\end{equation}
The incident magnetic field is taken along x-axis so that
\begin{equation}
\label{eq10}
h_{1} \ne 0, h_{2} = 0 = h_{3} = h_{4} 
\end{equation}
The first set of Maxwell's equations  
\begin{equation}
\label{eq11}
F_{ij;k} + F_{jk,i} + F_{ki;j} = 0,
\end{equation}
lead to
\begin{equation}
\label{eq12}
F_{23} = \mbox{constant} = H \mbox{(say)}.
\end{equation}
The semicolon represents a covariant differentiation. Here $F_{12} = F_{24} = F_{34} = 0$ due 
to assumption of infinite electromagnetic conductivity. The only non-vanishing component of $F_{ij}$ 
is $F_{23}$.  

The Einstein's field equations (with $\frac{8\pi G}{c^{4}} = 1$) 
\begin{equation}
\label{eq13}
R^{j}_{i} - \frac{1}{2} R g^{j}_{i} + \Lambda g^{j}_{i}  = - T^{j}_{i},
\end{equation}
for the line-element (\ref{eq1}) lead to the following system of equations:  
\[
\frac{1}{A^{2}}\left[- \frac{B_{44}}{B} - \frac{C_{44}}{C} + \frac{A_{4}}{A}\left(\frac{B_{4}}{B} + 
\frac{C_{4}}{C}\right) + \frac{A_{1}}{A}\left(\frac{B_{1}}{B} + \frac{C_{1}}{C}\right) 
+ \frac{B_{1}C_{1}}{BC}  - \frac{B_{4} C_{4}}{B C} \right]
\]
\begin{equation}
\label{eq14}
= p - \lambda - \frac{H^{2}}{2\bar{\mu} B^{2} C^{2}} + \Lambda ,
\end{equation}
\begin{equation}
\label{eq15}
\frac{1}{A^{2}}\left[- \left(\frac{A_{4}}{A}\right)_{4} + \left(\frac{A_{1}}{A}\right)_{1} - \frac{C_{44}}{C} 
+ \frac{C_{11}}{ C} \right]=   p + \frac{H^{2}}{2\bar{\mu} B^{2} C^{2}} + \Lambda,
\end{equation}
\begin{equation}
\label{eq16}
\frac{1}{A^{2}}\left[- \left(\frac{A_{4}}{A}\right)_{4} + \left(\frac{A_{1}}{A}\right)_{1} - \frac{B_{44}}{B} 
+ \frac{B_{11}}{B}\right] =   p + \frac{H^{2}}{2\bar{\mu} B^{2} C^{2}} + \Lambda,
\end{equation}
\[
\frac{1}{A^{2}}\left[ - \frac{B_{11}}{B} - \frac{C_{11}}{C} + \frac{A_{1}}{A}\left(\frac{B_{1}}{B} + 
\frac{C_{1}}{C}\right) + \frac{A_{4}}{A}\left(\frac{B_{4}}{B} + \frac{C_{4}}{C}\right) -
\frac{B_{1}C_{1}}{BC}  + \frac{B_{4} C_{4}}{B C}\right]
\]
\begin{equation}
\label{eq17}
= \rho + \frac{H^{2}}{2\bar{\mu} B^{2} C^{2}} - \Lambda,
\end{equation}
\begin{equation}
\label{eq18}
\frac{B_{14}}{B} + \frac{C_{14}}{C} - \frac{A_{4}}{A}\left(\frac{B_{1}}{B} + \frac{C_{1}}{C}
\right) - \frac{A_{1}}{A}\left(\frac{B_{4}}{B} + \frac{C_{4}}{C}\right) = 0,
\end{equation}
where the sub indices $1$ and $4$ in A, B, C and elsewhere denote ordinary differentiation
with respect to $x$ and $t$ respectively.

The rotation $\omega^{2}$ is identically zero. The scalar expansion $\theta$, shear scalar 
$\sigma^{2}$, acceleration vector $\dot{u}_{i}$ and proper volume $V^{3}$ are respectively 
found to have the following expressions:
\begin{equation}
\label{eq19}
\theta = u^{i}_{;i} = \frac{1}{A}\left(\frac{A_{4}}{A} + \frac{B_{4}}{B} + \frac{C_{4}}{C}
\right),
\end{equation}
\begin{equation}
\label{eq20}
\sigma^{2} = \frac{1}{2} \sigma_{ij} \sigma^{ij} = \frac{1}{3}\theta^{2} - \frac{1}{A^{2}}
\left(\frac{A_{4}B_{4}}{AB} + \frac{B_{4}C_{4}}{BC} + \frac{C_{4}A_{4}}{CA}\right),
\end{equation}
\begin{equation}
\label{eq21}
\dot{u}_{i} = u_{i;j}u^{j} = \left(\frac{A_{1}}{A}, 0, 0, 0\right) 
\end{equation}
\begin{equation}
\label{eq22}
V^{3} = \sqrt{-g} = A^{2} B C,
\end{equation}
where $g$ is the determinant of the metric (\ref{eq1}). 
%%%%%%%%%%%%%%%%%%%%%%%%%%%%%%%%%%%%%%%%%%%%%%%%%%%%%%%%%%%%%%%%%%%%%%%%%%%%%%%%%%%
%%%%%%%%%%%%%%%%%%%%%%%%%%%%%%%  SECTION 3  %%%%%%%%%%%%%%%%%%%%%%%%%%%%%%%%%%%%%%%%
\section{Solutions of the Field  Equations}

As in the case of general-relativistic cosmologies, the introduction of inhomogeneities 
into the string cosmological equations produces a considerable increase in mathematical 
difficulty: non-linear partial differential equations must now be solved. In practice, 
this means that we must proceed either by means of approximations which render the non-
linearities tractable, or we must introduce particular symmetries into the metric of the 
space-time in order to reduce the number of degrees of freedom which the inhomogeneities 
can exploit. \\

Here to get a determinate solution, let us assume that expansion ($\theta$) 
in the model is proportional to the eigen value $\sigma^{1}~~_{1}$ of  the shear tensor 
$\sigma^{i}~~_{j}$. This condition leads to
\begin{equation}
\label{eq23}
A = (BC)^{n},
\end{equation}
where $n$ is a constant. Equations (\ref{eq15}) and (\ref{eq16}) lead to
\begin{equation}
\label{eq24}
\frac{B_{44}}{B} - \frac{B_{11}}{B} = \frac{C_{44}}{C} - \frac{C_{11}}{C}.
\end{equation}
Using (\ref{eq23}) in (\ref{eq18}), yields 
\begin{equation}
\label{eq25}
\frac{B_{41}}{B} + \frac{C_{41}}{C} - 2n \left(\frac{B_{4}}{B} + 
\frac{C_{4}}{C}\right)\left(\frac{B_{1}}{B} + \frac{C_{1}}{C}\right) = 0.
\end{equation}
To find out deterministic solutions, we consider
\begin{equation}
\label{eq26}  
B = f(x)g(t) ~ ~ \mbox{and} ~ ~ C = h(x) k(t).
\end{equation}
In this case equation (\ref{eq25}) reduces to
\begin{equation}
\label{eq27}
\frac{f_{1}/f}{h_{1}/h} = - \frac{(2n - 1)(k_{4}/k) + 2n(g_{4}/g)}{(2n - 1)(g_{4}/g) + 
2n(k_{4}/k)} = K \mbox{(constant)},
\end{equation}
which leads to
\begin{equation}
\label{eq28}
\frac{f_{1}}{f} = K\frac{h_{1}}{h}
\end{equation}
and
\begin{equation}
\label{eq29}
\frac{k_{4}/k}{g_{4}/g} = \frac{K - 2nK - 2n}{2nK + 2n - 1} = a \mbox{(constant)}.
\end{equation}
From Eqs. (\ref{eq28}) and (\ref{eq29}), we obtain
\begin{equation}
\label{eq30}
f = \alpha h^{K}
\end{equation}
and
\begin{equation}
\label{eq31}
k = \delta g^{a},
\end{equation}
where $\alpha$ and $\delta$ are integrating constants. Eq. (\ref{eq24}) and (\ref{eq26}) 
reduce to  
\begin{equation}
\label{eq32}
\frac{g_{44}}{g} - \frac{k_{44}}{k} = \frac{f_{11}}{f} - \frac{h_{11}}{h} = N,
\end{equation}
where $N$ is a constant. Eqs. (\ref{eq29}) and (\ref{eq32}) 
lead to   
\begin{equation}
\label{eq33}
g g_{44} + a g_{4}^{2} = - \frac{N}{a - 1}g^{2},
\end{equation}
which leads to 
\begin{equation}
\label{eq34}
g = \beta^{\frac{1}{a + 1}}\sinh^{\frac{1}{a + 1}}(bt + t_{0}),
\end{equation}
where $\beta$ and $t_{0}$ are constants of integration and 
$$b = \sqrt{\frac{N(a + 1)}{1 - a}}. $$
Thus from Eq. (\ref{eq31}) we get
\begin{equation}
\label{eq35}
k = \delta \beta^{\frac{a}{a + 1}}\sinh^{\frac{a}{a + 1}}(bt + t_{0}).
\end{equation}
From Eqs. (\ref{eq27}) and (\ref{eq32}), we obtain 
\begin{equation}
\label{eq36}
h h_{11} + K h_{1}^{2} = \frac{N}{K - 1}h^{2},
\end{equation}
which leads to
\begin{equation}
\label{eq37}
h = \ell^{\frac{1}{K + 1}}\sinh^{\frac{1}{K + 1}}(rx + x_{0}),
\end{equation}
where $\ell$ and $x_{0}$ are constants of integration and 
$$ r = \sqrt{\frac{N(K + 1)}{K - 1}}.$$
Hence from Eq. (\ref{eq30}) we have 
\begin{equation}
\label{eq38}
f = \alpha \ell^{\frac{K}{K + 1}}\sinh^{\frac{K}{K + 1}}(rx + x_{0}).
\end{equation}
It is worth mentioned here that equations (\ref{eq33}) and (\ref{eq36}) are 
fundamental basic differential equations for which we have reported new 
solutions given by equations (\ref{eq34}) and (\ref{eq37}). \\
  
Thus, we obtain 
\begin{equation}
\label{eq39}
B = fg = Q\sinh^{\frac{K}{K + 1}}(rx + x_{0})\sinh^{\frac{1}{a + 1}}(bt + t_{0}),
\end{equation}
\begin{equation}
\label{eq40}
C = hk = R\sinh^{\frac{1}{K + 1}}(rx + x_{0})\sinh^{\frac{a}{a + 1}}(bt + t_{0}),
\end{equation}
and
\begin{equation}
\label{eq41}
A = (BC)^{n} = M\sinh^{n}(rx + x_{0})\sinh^{n}(bt + t_{0}),
\end{equation}
where
$$ Q = \alpha \beta^{\frac{1}{a + 1}}\ell^{\frac{K}{K + 1}},$$
$$ R = \delta \beta^{\frac{a}{a + 1}}\ell^{\frac{1}{K + 1}},$$
$$ M = (QR)^{n}.$$
Hence the geometry of the space-time (\ref{eq1}) takes the form
\[
ds^{2}= M^{2}\sinh^{2n}(rx + x_{0})\sinh^{2n}(bt + t_{0}) (dx^{2} - dt^{2}) + 
\]
\[
Q^{2} \sinh^{\frac{2K}{K + 1}}(rx + x_{0})\sinh^{\frac{2}{a + 1}}(bt + t_{0})dy^{2} + 
\]
\begin{equation}
\label{eq42}
R^{2} \sinh^{\frac{2}{K + 1}}(rx + x_{0})\sinh^{\frac{2a}{a + 1}}(bt + t_{0})dz^{2}. 
\end{equation}
By using the following transformation
\[
rX = rx + x_{0},
\]
\[
Y = Q y,
\]
\[
Z = R z
\]
\begin{equation}
\label{eq43}
bT = bt + t_{0} 
\end{equation}
the metric (\ref{eq42}) reduces to
\[
ds^{2}= M^{2}\sinh^{2n}(r X) \sinh^{2n}(b T) (dX^{2} - dT^{2}) + 
\]
\begin{equation}
\label{eq44}
\sinh^{\frac{2K}{K + 1}}(r X)\sinh^{\frac{2}{a + 1}}(b T)dY^{2} +  
\sinh^{\frac{2}{K + 1}}(r X)\sinh^{\frac{2a}{a + 1}}(b T)dZ^{2}. 
\end{equation}
%%%%%%%%%%%%%%%%%%%%%%%%%%%%%%%%%%%%%%%%%%%%%%%%%%%%%%%%%%%%%%%%%%%%%%%%%%%%%%%%%%%
%%%%%%%%%%%%%%%%%%%%%%%%%%%%%%%  SECTION 4  %%%%%%%%%%%%%%%%%%%%%%%%%%%%%%%%%%%%%%%%
\section{Some Physical and Geometric Properties of the Model}

In this case the physical parameters, i.e. the pressure $(p)$, the energy density $(\rho)$, the 
string tension density $(\lambda)$, the particle density $(\rho_{p})$ and the cosmological term 
$\Lambda(t)$ for the model (\ref{eq42}) are given by  
\[
p = \frac{1}{M^{2}\sinh^{2n}(bT) \sinh^{2n}(rX)}\Biggl[b^{2}\left\{n + \frac{a}{(a + 1)^{2}}\right\}\coth^{2}(bT)
\]
\[
- \,r^{2}\left\{n + \frac{K}{(K + 1)^{2}}\right\}\coth^{2}(rX) - b^{2}\left\{n + \frac{1}{(a + 1)}\right\} 
+ r^{2}\left\{n + \frac{K}{(K + 1)}\right\}\Biggr] 
\]
\begin{equation}
\label{eq45}
- \, \frac{\kappa}{\sinh^{2}(bT) \sinh^{2}(r X)} - \Lambda,
\end{equation}
\[
\lambda = \frac{1}{M^{2}\sinh^{2n}(bT) \sinh^{2n}(rX)}\Biggl[r^{2}\left\{n + \frac{K}{(K + 1)}\right\} 
- \, b^{2}\left\{n + 1 + \frac{1}{(a + 1)}\right\}
\]
\begin{equation}
\label{eq46}
- \, 2r^{2}\left\{n + \frac{K}{(K + 1)^{2}}\right\}\coth^{2}(rX)\Biggr] - \, \frac{2\kappa}{\sinh^{2}(bT) 
\sinh^{2}(r X)},
\end{equation}
\[
\rho = \frac{1}{M^{2}\sinh^{2n}(bT) \sinh^{2n}(rX)}\Biggl[b^{2}\left\{n + \frac{a}{(a + 1)^{2}}\right\}
\coth^{2}(bT)
\]
\begin{equation}
\label{eq47}
+ \,r^{2}\left\{n + \frac{K}{(K + 1)^{2}}\right\}\coth^{2}(rX) - r^{2} \Biggr] 
- \, \frac{\kappa}{\sinh^{2}(bT) \sinh^{2}(r X)} + \Lambda,
\end{equation}
\[
\rho_{p} =  \rho - \lambda = \frac{1}{M^{2}\sinh^{2n}(bT) \sinh^{2n}(rX)}\Biggl[b^{2}
\left\{n + \frac{a}{(a + 1)^{2}}\right\}\coth^{2}(bT)
\]
\[
+ \,3r^{2}\left\{n + \frac{K}{(K + 1)^{2}}\right\}\coth^{2}(rX) + b^{2}\left\{n - 1 + \frac{1}{(a + 1)}
\right\} 
\]
\begin{equation}
\label{eq48}
- \, r^{2}\left\{n + 1 + \frac{K}{(K + 1)}\right\}\Biggr] + \, \frac{\kappa}{\sinh^{2}(bT) \sinh^{2}(r X)} + \Lambda,
\end{equation}
where
$$ \kappa = \frac{H^{2}}{2\bar{\mu}}.$$
For the specification of $\Lambda$, we assume that the fluid obeys an equation of state of the form
\begin{equation}
\label{eq49}
p = \gamma \rho,
\end{equation}
where $\gamma(0 \leq \gamma \leq 1)$ is a constant. \\
From Eqs. (\ref{eq45}), (\ref{eq47}) and (\ref{eq49}), we obtain
\[
\Lambda = \frac{1}{(1 - \gamma) M^{2}\sinh^{2n}(bT) \sinh^{2n}(rX)}\Biggl[(1 - \gamma) b^{2}\left\{n + \frac{a}
{(a + 1)^{2}}\right\}\coth^{2}(bT)
\]
\[
- \,(1 + \gamma)r^{2}\left\{n + \frac{K}{(K + 1)^{2}}\right\}\coth^{2}(rX) - b^{2}\left\{n + \frac{1}{(a + 1)}
\right\}
\]
\begin{equation}
\label{eq50}
 + \, r^{2}\left\{n + \frac{K}{(K + 1)}\right\} - \gamma r\Biggr] - \, \frac{\kappa}{\sinh^{2}(bT) 
\sinh^{2}(r X)}.
\end{equation}
%%%%%%%%%%%%%%%%%%% Figure 1 %%%%%%%%%%%%
\begin{figure}[htbp]
\centering
\includegraphics[width=8cm,height=8cm,angle=0]{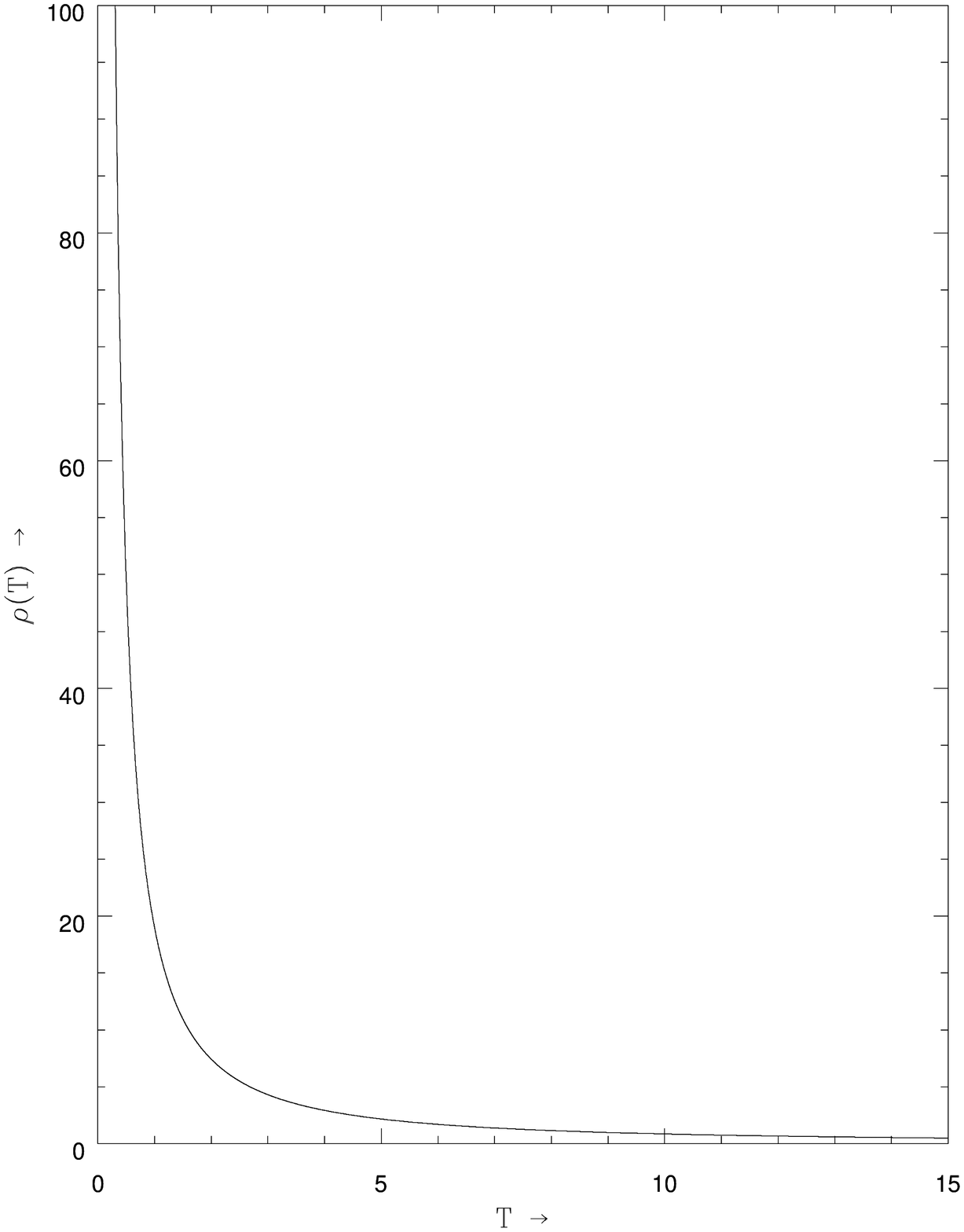}
\caption{The plot of energy density $\rho(T)$ Vs. time}
\end{figure}
%%%%%%%%%%%%%%%%%%%%%%%%%%%%%%%%%% %%%%%%%%
%%%%%%%%%%%%%%%%%%% Figure 2 %%%%%%%%%%%%
\begin{figure}[htbp]
\centering
\includegraphics[width=8cm,height=8cm,angle=0]{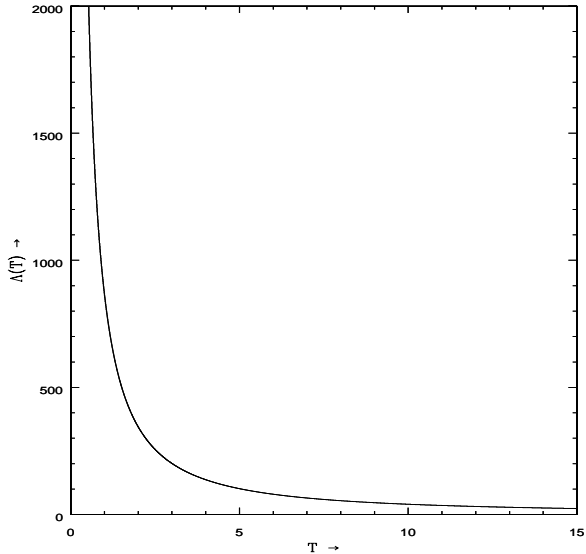}
\caption{The plot of cosmological term $\Lambda(T)$ Vs. time}
\end{figure}
%%%%%%%%%%%%%%%%%%%%%%%%%%%%%%%%%%%%%%%%%%%%%%%%
From Eq. (\ref{eq47}), we note that $\rho(t)$ is a decreasing function of time and $\rho > 0$ for 
all times. Figure 1 shows this behaviour of energy density.\\ 

In spite of homogeneity at large scale our universe is inhomogeneous at small scales, so physical 
quantities being position dependent are more natural in our observable universe if we do not go 
to super high scale. This result shows this kind of physical importance. In recent time the $\Lambda$-term 
has interested theoreticians and observers for various reasons. The nontrivial role of the vacuum in 
the early universe generate a $\Lambda$-term that leads to inflationary phase. Observationally, this term 
provides an additional parameter to accommodate conflicting data on the values of the Hubble constant, the 
deceleration parameter, the density parameter and the age of the universe (for example, see the references 
\cite{ref62,ref63}). Assuming that $\Lambda$ owes its origin to vacuum interactions, as suggested in 
particular by Sakharov \cite{ref64}, it follows that it would in general be a function of space and time 
coordinates, rather than a strict constant. In a homogeneous universe $\Lambda$ will be at most time dependent 
\cite{ref65}. In our case this approach can generate $\Lambda$ that varies both with space and time. In 
considering the nature of local massive objects, however, the space dependence of $\Lambda$ cannot be ignored. 
For details discussion, the readers are advised to see the references (Narlikar, Pecker and Vigier 
\cite{ref66}, Ray and Ray \cite{ref67}, Tiwari, Ray and Bhadra \cite{ref68}).  \\ 

The behaviour of the universe in this model will be determined by the cosmological 
term $\Lambda$ ; this term has the same effect as a uniform mass density $\rho_{eff} 
= - \Lambda / 4\pi G$, which is constant in space and time. A positive value of 
$\Lambda$ corresponds to a negative effective mass density (repulsion). Hence, we 
expect that in the universe with a positive value of $\Lambda$, the expansion will 
tend to accelerate; whereas in the universe with negative value of $\Lambda$, 
the expansion will slow down, stop and reverse. From Eq. (\ref{eq50}), we see 
that the cosmological term $\Lambda$ is a decreasing function of time and it 
approaches a small positive value as time increases more and more. From Figure 2 we note this behaviour 
of cosmological term $\Lambda$. Recent cosmological 
observations (Garnavich et al. \cite{ref54}, Perlmutter et al. \cite{ref51}, 
Riess et al. \cite{ref52,ref57}, Schmidt et al. \cite{ref56}) suggest the 
existence of a positive cosmological constant $\Lambda$ with the magnitude 
$\Lambda(G\hbar/c^{3})\approx 10^{-123}$. These observations on magnitude and 
red-shift of type Ia supernova suggest that our universe may be an accelerating 
one with induced cosmological density through the cosmological $\Lambda$-term. 
Thus, our model is consistent with the results of recent observations. \\  

The kinematical quantities , i.e. the scalar of expansion $(\theta)$, 
shear tensor $(\sigma)$, the acceleration vector $(\dot{u}_{i})$ and the proper 
volume $(V^{3})$ for the model (\ref{eq42}) are given by 
\begin{equation}
\label{eq51}
\theta = \frac{b(n + 1) \coth(bT)}{M \sinh^{n}(bT) \sinh^{n}(rX)},
\end{equation}
\begin{equation}
\label{eq52}
\sigma^{2} = \frac{b^{2} \coth^{2}(bT)[(a + 1)^{2}(n^{2} - n + 1) - 3a]}{3(a + 1)^{2} M^{2}\sinh^{2n}(bT) 
\sinh^{2n}(rX)},
\end{equation}
\begin{equation}
\label{eq53}
\dot{u}_{i} = (n r \coth(rX), 0, 0, 0),
\end{equation}
\begin{equation}
\label{eq54}
V^{3} = \sinh^{2n + 1}(bT) \sinh^{2n + 1}(rX).
\end{equation}
From Eqs. (\ref{eq47}) and (\ref{eq48}), we obtain
\begin{equation}
\label{eq55}
\frac{\sigma^{2}}{\theta^{2}} = \frac{(a + 1)^{2}(n^{2} - n + 1) - 3a}{3(n + 1)^{2}(a + 1)^{2}} = \mbox {constant}.
\end{equation}
The dominant energy conditions (Hawking and Ellis \cite{ref69})
$$
(i) ~ ~ \rho - p \geq 0 ~ ~ ~ (ii) ~ ~ \rho + p \geq 0 $$
lead to
\[
2r^{2}\left\{n + \frac{K}{(K + 1)^{2}}\right\} \coth^{2}(rX) - r^{2}\left\{n + 1 +\frac{K}{K + 1}\right\}
\]
\begin{equation}
\label{eq56}
+ \, b^{2}\left\{n + \frac{1}{a + 1}\right\} + 2\Lambda M^{2} \sinh^{2n}(bT) \sinh^{2n}(rX) \geq 0,
\end{equation}
and
\[
2b^{2}\left\{n + \frac{a}{(a + 1)^{2}}\right\} \coth^{2}(bT) + r^{2}\left\{n - 1 +\frac{K}{K + 1}\right\}
\]
\begin{equation}
\label{eq57}
- \, b^{2}\left\{n + \frac{1}{a + 1}\right\} \geq  2 M^{2} \kappa  \sinh^{2n- 2}(bT) \sinh^{2n - 2}(rX). 
\end{equation}
The reality conditions (Ellis \cite{ref70})
$$ (i)~ ~ \rho + p > 0, ~ ~ ~ (ii) ~ ~ \rho + 3p > 0, $$
lead to
\[
2b^{2}\left\{n + \frac{a}{(a + 1)^{2}}\right\} \coth^{2}(bT) + r^{2}\left\{n - 1 +\frac{K}{K + 1}\right\}
\]
\begin{equation}
\label{eq58}
- \, b^{2}\left\{n + \frac{1}{a + 1}\right\} >  2 M^{2} \kappa  \sinh^{2n- 2}(bT) \sinh^{2n - 2}(rX), 
\end{equation}
and
\[
4b^{2}\left\{n + \frac{a}{(a + 1)^{2}}\right\} \coth^{2}(bT) - 2r^{2}\left\{n +\frac{K}{(K + 1)^{2}}\right\}
\coth^{2}(rX)
\]
\[
+ 3r^{2}\left\{n +\frac{K}{K + 1}\right\} - \,3 b^{2}\left\{n + \frac{1}{a + 1}\right\} -r^{2}
\]
\begin{equation}
\label{eq59}
> 4 M^{2}\kappa \sinh^{2n- 2}(bT) \sinh^{2n - 2}(rX) + 2 M^{2}\Lambda \sinh^{2n}(bT) \sinh^{2n}(rX)  . 
\end{equation}
The model starts expanding with a big bang at $T = 0$ and it stops expanding at $T = \infty$. In general, 
the model represents an expanding, shearing and non-rotating universe. Since $\frac{\sigma}{\theta} = $ 
constant, hence the model does not approach isotropy. When the uniform magnetic field is not present and 
$p = 0$, $\Lambda = 0$, our solution represents the solution obtained by Baysal et al. \cite{ref59}. The 
model is accelerating. The proper volume in the model increases as $T$ increases. \\ 
%%%%%%%%%%%%%%%%%%%%%%%%%%%%%%%%%%%%%%%%%%%%%%%%%%%%%%%%%%%%%%%%%%%%%%%%%%%%%%%%%%%
%%%%%%%%%%%%%%%%%%%%%%%%%%%%%%%  SECTION 5  %%%%%%%%%%%%%%%%%%%%%%%%%%%%%%%%%%%%%%%%
\section{Solutions of the Field  Equations in Absence of Magnetic Field}

In absence of the magnetic field i.e. $H = 0$, the physical parameters, i.e. the pressure $(p)$, the 
energy density $(\rho)$, the string tension density $(\lambda)$, the particle density $(\rho_{p})$ and 
the cosmological term $\Lambda(t)$ for the model (\ref{eq44}) are given by  
\[
p = \frac{1}{M^{2}\sinh^{2n}(bT) \sinh^{2n}(rX)}\Biggl[b^{2}\left\{n + \frac{a}{(a + 1)^{2}}\right\}\coth^{2}(bT)
\]
\begin{equation}
\label{eq60}
- \,r^{2}\left\{n + \frac{K}{(K + 1)^{2}}\right\}\coth^{2}(rX) - b^{2}\left\{n + \frac{1}{(a + 1)}\right\} 
+ r^{2}\left\{n + \frac{K}{(K + 1)}\right\}\Biggr] - \Lambda, 
\end{equation}
\[
\lambda = \frac{1}{M^{2}\sinh^{2n}(bT) \sinh^{2n}(rX)}\Biggl[r^{2}\left\{n + \frac{K}{(K + 1)}\right\} 
- \, b^{2}\left\{n + 1 + \frac{1}{(a + 1)}\right\}
\]
\begin{equation}
\label{eq61}
- \, 2r^{2}\left\{n + \frac{K}{(K + 1)^{2}}\right\}\coth^{2}(rX)\Biggr],
\end{equation}
\[
\rho = \frac{1}{M^{2}\sinh^{2n}(bT) \sinh^{2n}(rX)}\Biggl[b^{2}\left\{n + \frac{a}{(a + 1)^{2}}\right\}
\coth^{2}(bT)
\]
\begin{equation}
\label{eq62}
+ \,r^{2}\left\{n + \frac{K}{(K + 1)^{2}}\right\}\coth^{2}(rX) - r^{2} \Biggr] + \Lambda,
\end{equation}
\[
\rho_{p} =  \rho - \lambda = \frac{1}{M^{2}\sinh^{2n}(bT) \sinh^{2n}(rX)}\Biggl[b^{2}
\left\{n + \frac{a}{(a + 1)^{2}}\right\}\coth^{2}(bT)
\]
\[
+ \,3r^{2}\left\{n + \frac{K}{(K + 1)^{2}}\right\}\coth^{2}(rX) + b^{2}\left\{n - 1 + \frac{1}{(a + 1)}
\right\} 
\]
\begin{equation}
\label{eq63}
- \, r^{2}\left\{n + 1 + \frac{K}{(K + 1)}\right\}\Biggr] + \Lambda.
\end{equation}
By using the equation of state (\ref{eq49}) in  Eqs. (\ref{eq60}) and (\ref{eq62}), we obtain
\[
\Lambda = \frac{1}{(1 - \gamma) M^{2}\sinh^{2n}(bT) \sinh^{2n}(rX)}\Biggl[(1 - \gamma) b^{2}\left\{n + \frac{a}
{(a + 1)^{2}}\right\}\coth^{2}(bT)
\]
\[
- \,(1 + \gamma)r^{2}\left\{n + \frac{K}{(K + 1)^{2}}\right\}\coth^{2}(rX) - b^{2}\left\{n + \frac{1}{(a + 1)}
\right\}
\]
\begin{equation}
\label{eq64}
 + \, r^{2}\left\{n + \frac{K}{(K + 1)}\right\} - \gamma r\Biggr].
\end{equation}
We observe that in absence of the magnetic field, the expressions for Kinematical quantities for the 
model (\ref{eq42}) are unchanged. 
%%%%%%%%%%%%%%%%%%% Figure 3 %%%%%%%%%%%%
\begin{figure}[htbp]
\centering
\includegraphics[width=8cm,height=8cm,angle=0]{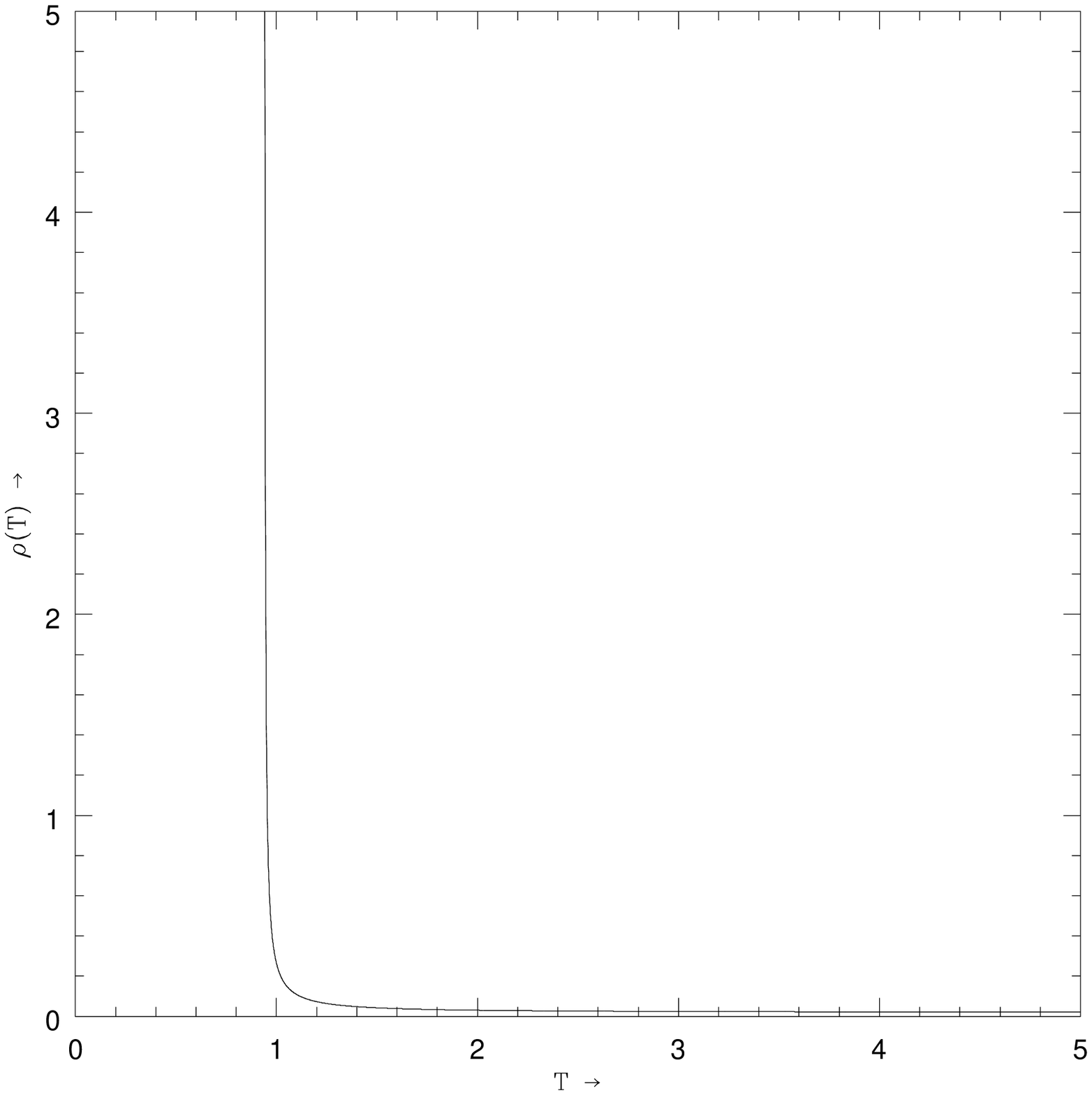}
\caption{The plot of energy density $\rho(T)$ Vs. time}
\end{figure}
%%%%%%%%%%%%%%%%%%%%%%%%%%%%%%%%%% %%%%%%%%
%%%%%%%%%%%%%%%%%%% Figure 4 %%%%%%%%%%%%
\begin{figure}[htbp]
\centering
\includegraphics[width=8cm,height=8cm,angle=0]{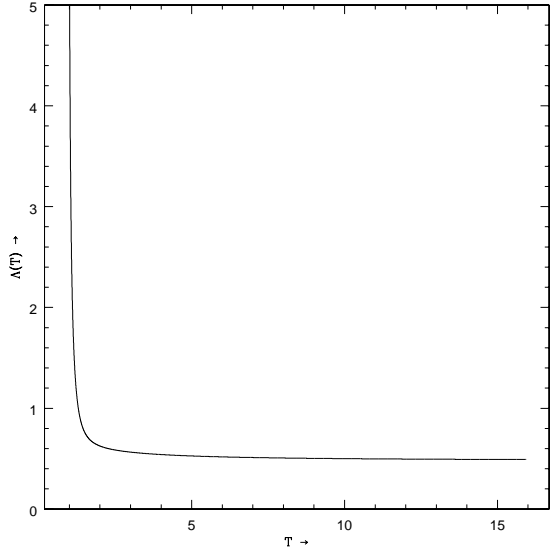}
\caption{The plot of cosmological term $\Lambda(T)$ Vs. time}
\end{figure}
%%%%%%%%%%%%%%%%%%%%%%%%%%%%%%%%%% %%%%%%%%
From Eq. (\ref{eq62}), we note that $\rho(t)$ is a decreasing function of time and $\rho > 0$ for 
all times. Figure 3 shows this behaviour of energy density. From Eq. (\ref{eq64}), we see 
that the cosmological term $\Lambda$ is a decreasing function of time and it 
approaches a small positive value as time increases more and more which matches with recent 
observations. From Figure 4 we note this behaviour of cosmological term $\Lambda$. When we set 
$p = 0$ and $\Lambda = 0$, our solution represents the solution obtained by Baysal et al. \cite{ref59}). 
%%%%%%%%%%%%%%%%%%%%%%%%%%%%%%%%%%%%%%%%%%%%%%%%%%%%%%%%%%%%%%%%%%%%%%%%%%%%%%%%%%%
%%%%%%%%%%%%%%%%%%%%%%%%%%%%%%%  SECTION 5  %%%%%%%%%%%%%%%%%%%%%%%%%%%%%%%%%%%%%%%%
\section{Concluding Remarks}
We have obtained a new cylindrically symmetric inhomogeneous cosmological model of uniform 
electromagnetic perfect fluid as the source of matter where the cosmological constant is varying 
with time. Generally the model represents an expanding, shearing and non-rotating universe in which 
the flow vector is geodetic. The model does not approach isotropy. \\

In presence and absence of magnetic field, the cosmological terms in the models are decreasing function 
of time and approach a small value  at late time. The values of cosmological ``constant'' for the models 
are found to be small and positive, which is supported by the results from supernovae observations 
recently obtained by the High-Z Supernovae Ia Team and Supernova Cosmological Project (Garnavich et al. 
\cite{ref54}, Perlmutter et al. \cite{ref51}, Riess et al. \cite{ref52,ref57}, Schmidt et al. \cite{ref56}).  
Our solutions generalize the solutions obtained by Baysal et al. \cite{ref59}).
\section*{Acknowledgements} 
Author would like to thank the Inter-University Centre for Astronomy and Astrophysics, 
Pune, India for providing facility under associateship programme where this work was 
carried out. 
%\newline
%\nonumsection{References}

\end{document}